\documentclass[prb,preprint]{revtex4-1} 

\usepackage{amsmath} 

\usepackage{graphicx} 

 \newcommand{\nd}{Nd\-Ba$_{2}$\-Cu$_{3}$\-O$_{7-\delta}$}
 \newcommand{\Sm}{Sm\-Ba$_{2}$\-Cu$_{3}$\-O$_{7-\delta}$}
 
 \newcommand{\NEG}{(Nd$_{0.33}$\-Eu$_{0.33}$\-Gd$_{0.33}$)\-Ba$_{2}$\-Cu$_{3}$\-O$_{y}$}
 
 \newcommand{\SEG}{(Sm$_{0.33}$\-Eu$_{0.33}$\-Gd$_{0.33}$)\-Ba$_{2}$\-Cu$_{3}$\-O$_{y}$}
 
 \newcommand{\y}{Y\-Ba$_{2}$\-Cu$_{3}$\-O$_{7-\delta}$}

 \newcommand{\Gd}{Gd\-Ba$_{2}$\-Cu$_{3}$\-O$_{7-\delta}$}
 
 \newcommand{\bizwei}{Bi$_{2}$\-Sr$_{2}$\-Ca\-Cu$_{2}$\-O$_{8+\delta}$}
 
\begin{document}

\title{Pinning force scaling analysis of Fe-based high-$T_c$ superconductors}

\author{M. R. Koblischka}
\affiliation{Experimental Physics, Saarland University, Campus, Building C 6 3,
   D-66123 Saarbr\"ucken, Germany.}
   
\author{M. Muralidhar}
\affiliation{Superconducting Materials Laboratory, Department of Materials Science and 
Engineering, Shibaura Institute of Technology, 3-7-5 Toyosu, Koto-ku, Tokyo 135-8548, Japan.}
\date{\today}

\begin{abstract}
Pinning force data, $F_{\rm p}$, of a variety of Fe-based high-$T_c$ superconductors
(11-, 111-, 122- and 1111-type) were analyzed by means of a scaling approach
based on own experimental data and an extensive collection of literature data.
The literature data were mostly replotted, but also converted from critical
current measurements together with data for the irreversibility line when available
from the same authors. Using the scaling approaches of Dew-Hughes [1]
and Kramer [2], we determined the scaling behavior and the best
fits to the theory. The data of most experiments analyzed show a good
scaling behavior at high temperatures when plotting the normalized pinning force
$F_{\rm p}/F_{\rm p,max}$
versus the irreversibility field, $H_{\rm irr}$. The resulting peak positions,
$h_0$, were found at $\approx$0.3 for the 11-type materials, at $\approx$0.48 for the
111-type materials, between 0.32 and 0.5 for the 1111-type materials
and between 0.25 and 0.71 for the 122-type materials. Compared to the
typical results of \bizwei\ ($h_0 \approx$0.22) and \y\ ($h_0 \approx$0.33), most of
the 122 and 1111 samples investigated show peak values higher than
0.4, which is similar to the data obtained on the light-rare earth
123-type HTSC like NdBa$_2$Cu$_3$O$_y$. This high peak position ensures a
good performance of the materials in high applied magnetic fields
and is, therefore, a very promising result concerning the possible
applications of the Fe-based high-$T_c$ superconductors. \\[0.7cm]
\noindent
[1] E. J. Kramer, J. Appl. Phys. {\bf 44}, 1360 (1973).\\[-0.3cm]
[2] D. Dew-Hughes, Philos. Mag. {\bf 30}, 293 (1974).

\end{abstract}

\maketitle

\section{Introduction}

The development of the iron-based pnictide \cite{1} and selenide \cite{2}
high-$T_c$ superconductors
has stimulated a vast research effort on these materials, even though their
transition temperatures are still not as high as the cuprate materials. From
the viewpoint of possible applications, the superconducting transition temperatures
of the pnictide materials are more similar to MgB$_2$ with a $T_c$ of around 40 K, but
the pnictides are also of ceramic nature
and share many features with the cuprates like the extremely high values for the upper
critical field, $H_{c2}$ \cite{Gurevich,Senatore}.
In the meantime, there are several Fe-based high-$T_c$ superconductors known which can be
classified into four main families (11-, 111-, 122- and 1111-type); the 122-family can
also be made up from Fe-pnictides as well as from selenides.
Most research on these materials is devoted to clarify the superconducting
nature, offering a different approach as the cuprates. In this field, several
reviews were already given in the literature \cite{Gurevich,Kamihara,Hoffman,Eskildsen}.

Concerning possible applications of these materials, the grain boundary
problem also plays an important issue for the Fe-based superconductors as discussed in Refs.
\cite{Durrell,Weiss}.
Due to this reason, the preparation of epitaxial thin films seem to be
the most interesting direction. However, an important general property of
a superconducting material is the flux pinning behavior, as flux pinning
will rule the achievable critical current densities, the position of the
irreversibility line and hence, effects of flux motion and creep. In order
to perform such studies, homogeneous materials with a high content of the
superconducting phase are required. This demand can be fulfilled with single
crystals of the respective phase, but also with epitaxially grown thin films
or phase-pure polycrystalline samples. Therefore, some research efforts had
to be invested to obtain materials of such quality \cite{Iida,Ghorbani,45,Inosov,Ishibashi,HLei}.

The best tool to study details of the underlying flux pinning mechanism(s)
is the scaling behavior of the flux pinning forces, $F_{\rm p} = j_c \times B$, determined
from the critical current densities, $j_c$. Such scaling was found to be useful
already on the conventional superconductors in the works of Kramer \cite{Kramer} and
Dew-Hughes (DH) \cite{DH}. A scaling of $F_{\rm p}$ was obtained when plotting the
normalized pinning force $F_{\rm p}/F_{\rm p,max}$ versus the reduced
field $h=H_a/H_{c2}$, where $H_{c2}$ denotes the upper critical field.
This scaling implies $F_{\rm p}=H_{c2}(T)^m \cdot f(h)^n$ with $m$ and $n$ being numerical
parameters describing the actual pinning mechanism; $f(h)$ depends only
on the reduced magnetic field, $h$. In the literature, several pinning functions
$f(h)$ are described depending on the size and character of the defects
providing the pinning based on the study of conventional conventional,
hard type-II superconductors \cite{Antesberger,CE,Fietz,Adaktylos}. The scaled pinning
force data were then fitted to the functional
dependence given by
\begin{equation}
F_{\rm p}/F_{\rm p,max} = A (h)^p (1-h)^q
\end{equation}
with $A$ being a numerical parameter, and $p$ and $q$ are describing the
actual pinning mechanism. The position of the maximum in
the $F_{\rm p}$ plot, $h_0$, is given by $p/p+q$.
In the model of DH, six different pinning functions $f(h)$ describing the core pinning
using Eq. (1) are given. (1) $p=$0, $q=$2: normal, volume pinning;
(2) $p=$1, $q=$ 1: $\Delta\kappa$-pinning, volume pins;
(3) $p=$1/2, $q=$2: normal, surface pins;
(4) $p=$3/2, $q=$1: $\Delta\kappa$-pinning, surface pins;
(5) $p=$1, $q=$2: normal, point pins; and
(6) $p=$2, $q=$1: $\Delta\kappa$-pinning, point pins.
Additionally, (3) is predicted by Kramer \cite{Kramer} for shear-breaking
in the case of a set of planar pins. The $\Delta\kappa$-pinning is nowadays called 
$\delta T_c$-pinning \cite{Blatter}. In a recent work \cite{Ma}, these six functions
plus the ones for magnetic pinning were analyzed and it was found that they are linearly
dependent, so some functions could be removed from the analysis. In the present case,
we regard only the  6 functions mentioned above.

For various high-$T_c$ cuprate materials, a
reasonably good scaling of $F_p$ is found as well, however, experiments
have shown that the appropriate scaling field is the irreversibility
field $H_{\rm irr}$ instead of $H_{c2}$. The use of the irreversibility field for
the scaling was already discussed in Refs. \cite{Fabbricatore,MK_PhysC,MK_PRB};
$H_{\rm irr}$ represents
the upper limit of strong flux pinning, not $H_{c2}$ as in the case of the
conventional superconductors, as by definition $F_{\rm p} \rightarrow 0$ at $H=H_{\rm irr}$.
In general, one can state that any determination of the parameters
$p$ and $q$ from scaling laws is more significant than one obtained only
from measurements of the irreversibility line.

Good pinning force scalings were reported in \y\ (Y-123), the light-rare-earth (LRE)-123 systems
like \nd\ (Nd-123), \Gd\ (Gd-123), \Sm\ (Sm-123)
and in the ternary LRE-compounds like \NEG\ (NEG) and \SEG\ (SEG) \cite{MK_JAP,21,14,15,16,MK_APL}.
The peak position
of the scaling obtained ranges from 0.33 up to 0.5; the latter indicating
the presence of the $\delta T_c$-pinning. The temperature range covered is mainly
between 60 K and $T_c$, which is on one hand corresponding to the experimentally available
magnetic field range, and on the other hand also containing the most interesting features
like the fishtail peak. Furthermore, it was attempted in \cite{MK_PRB} to include the effects
of flux creep in the DH model, but it turned out that especially the peak position $h_0$ is
independent of creep effects. For more details, see the reviews given
in Refs. \cite{MK_PhysC,15,16}. The addition of nanoparticles to the superconducting
matrix of the NEG system may lead to an apparent non-scaling behavior as shown in Ref. \cite{MK_PSS,17}.
In the case of Bi-based cuprate superconductors, the scaling
leads to a peak position of $h_0 \approx$0.22, which indicates the dominance of
grain boundary pinning. In Bi-2223, a distinct non-scaling was observed
indicating the change of the dominating pinning mechanism with temperature;
the peak position shifts from 0.33 to 0.2 with increasing temperature \cite{MK_EPJ}.
The other cuprate materials fit in this basic scheme depending on the
degree of anisotropy, see the data in Refs. \cite{15,16,Welp}. Very recently, the
pinning force scaling were reviewed by Sandu \cite{Sandu}, also presenting the data
of MgB$_2$ and with various additions to increase the flux pinning.

In the case of the iron-based superconductors, which consist of four
main families, the pinning force scaling was obtained as well, also here
mainly in the high temperature range which is accessible to the experiments.
From the very beginning, one can see here a wider range of chemical dopings
which causes a larger variation of the pinning force scaling as compared to
the cuprate materials. Therefore, it is important to have a comparison of
the available experimental data in order to discuss the flux pinning
properties of these materials in detail.

In the present work, the present literature data of the flux pinning
scaling are collected together and compared to each other.
A total of 31 Fe-based superconducting
compositions was retrieved from the
literature \cite{22,23,24,25,26,27,28,29,30,31,32,33,x,xx,y,34,35,36,37,38,xxx,39,40,41,42,43,44}.
However, in order to achieve the goal of a valid
comparison of the data, several data sets of the literature had to be replotted and
reworked to enable such a comparison. Additionally, some own data
are presented.

\section{Experimental procedure and analysis details}

The flux pinning data are normally obtained via an elaboration of magnetic data
measured by VSM- or SQUID magnetometry. Additionally, also data from electric
transport measurements can be converted into flux pinning data. Here, it is
important to note that the criteria for determining the critical current
density and the irreversibility line must be the same for a given experiment.
The reduced field, $h=H_a/H_{\rm irr}$, is normally
determined from the irreversibility fields
according to a voltage criterion of 1 $\mu$Vcm$^{-1}$ (i.e., the same criterion as
applied for the $j_c$ determination). In case of magnetization data, the criterion
for determining $H_{\rm irr}$ is typically
chosen as 1 $\times$ 10$^4$ A/cm$^2$.

In several papers in the literature, the pinning force scaling is
performed using the peak field of a fishtail peak, $H_p$, or the field where
the maximum pinning force occurs, $H(F_{\rm p,max})$, instead of $H_{\rm irr}$.
$H_p$ or $H(F_{\rm p,max})$ are experimentally directly accessible, so several
authors prefer these fields, whereas the determination of $H_{\rm irr}$ requires
the use of a criterion, which can be arbitrarily fixed. However, the scaling is
not always properly done, so one has to carefully check the procedures applied.
Therefore, one has to re-plot the according data sets in
order to allow a direct comparison of the results. Some authors have only
presented data for the peak position, $h_0$, and not the full parameters of a
corresponding fit. These data sets were fitted by taking the parameters from the
respective master curve, and the fit was started using the basic parameters of the
DH model. The results of these treatments are indicated in Tables 1-4 by
using italics. In some cases, the pinning force data or critical current
densities are published without showing the results for $H_{\rm irr}$, so these
data sets had to be excluded from the present analysis.

\section{Results and discussion}

In the following, the results of the pinning force scaling are presented for each of the
four families of Fe-based superconductors.
In Tables 1-4, all the contributions are sorted according
to the publication year, and the type of sample (sc denotes single crystal, film, tape or
polycrystalline) and some comments to the scaling are given.
Figures 1-4 contain the determined pinning functions according to the parameters given in
Tables 1-4. To enable an easier comparision of all the data, the data of the single crystals (sc) 
are plotted using filled symbols, the data
of thin films and tape samples are plotted using open symbols. Data of different samples but
of the same author are drawn using the same symbols and color, but with different lines.

\subsection{122-family}
In the case of the 122-family (both pnictides and selenides), the largest number
of works was performed concerning the pinning force analysis. 
The 122-family shows some importance for possible
applications \cite{Gurevich,Weiss,Durrell}, so not only single crystals, but also
epitaxial thin films and IBAD tapes were investigated in the literature.
The peak positions determined range between 0.22
and 0.5, with a remarkable exception of the data by L. Fang {\em et al.} \cite{25}, which show
peaks in the pinning force scaling as high as 0.68. In this work, also effects
of irradiation on the flux pinning properties were investigated. The irradiated samples
show indeed higher peak positions, which indicates the importance of flux pinning provided
by a variation of the transition temperature, $T_c$.

Table 1 summarizes all pinning force scaling data published in the literature
so far; the resulting pinning functions are plotted in Fig. 1 in order to enable a
direct comparison of the data; the data set of the iron selenides (13) is 
indicated by a darker background in Table 1.\\

The available $F_p$-data show a clear tendency towards $h_0$ values higher than 0.33.
However, it is important to note here that the peak positions larger than 0.33 can
only be reached by an additional flux pinning contribution of the $\delta T_c$-type;
the flux pinning provided by normal
conducting particles and obstacles can only yield $h_0$-values up to 0.33.
Furthermore, it is remarkable that the data of thin films and IBAD samples (7,10-13, marked using open symbols) show
peak positions located at 0.22 to 0.33, which is also seen for films of other iron pnictide families.
This indicates that the pinning landscape created by the substrate provides a strong pinning
contribution, independent from the material properties.\\
The data of the single crystalline materials (1-6,8,9,13,15,17-19) are plotted in Fig. 1 using filled symbols.
The data (7) and (10) exhibit a typical non-scaling over the entire
temperature range -- a good scaling is only obtained at higher temperatures close to $T_c$ with a
high $h_0$, whereas at lower temperatures, the peak position is shifting towards smaller values.
This behavior indicates a change of the dominant flux pinning mechanism with increasing temperature.\\
The data (3), (4) and (5) are single crystals with different doping, where
each of the three crystals exhibits a distinct scaling behavior.\\
The data sets (2), (11), (15) and (19) are the ones with the highest peak position, $h_0$; while the data (12) show a
peak at 0.33, but an extremely broad scaling curve, which yields strong flux pinning in the high-field regime.
The data of the iron selenide (13) fit well to the other family members; the peak position, $h_0$, is found at
0.32.\\
The data of the ploycrystalline material (16) are not much different from
the single crystal data of the same composition (6). The pinning function (6) was also observed in
own SQUID-data of a single crystal, which are not shown here.

\subsection{1111-family}
In the 1111-family, only some investigations of the flux pinning force scaling are
reported; but also on various types of materials (single crystals, polycrystalline
samples and films). All pinning force scaling data are collected in Table 2 and Figure 2.
It is striking to see that the literature data do not follow any common behavior as all reported
pinning force scalings yield a different peak position, $h_0$.
Interestingly enough, practically all data with the exception of the films (3) point to $h_0$ values
larger than 0.33, which indicates a strong contribution of the $\delta T_c$-pinning in this type
of material. Also here, the data measured on the IBAD tape exhibit the lowest peak position, which
is, however, comparable to that of the 122-family data and coincides well with the Kramer theory.
The data set (1) exhibits an extremely broad and high peak in the scaling diagram \cite{35},
even though the sample measured is of the polycrystalline type. The resulting $h_0$ is 0.57, which
indicates a strong contribution of the $\delta T_c$-type.\\
The data (2) stem from the only single crystals measured in the literature \cite{36}.
The authors have plotted their data in a scaling versus $H(F_{p,max})$, but for their analysis they
correctly use an adapted version of the scaling law of the form $f(h) = h^p(2-h)^q$, which
accounts that the maximal value of $H(F_{\rm p,max})$ is 2 instead of 1 (their figure D3).
For this reason, their determined values of $q$ and $p$ correspond well with the DH theory,
and the peak position
of the $F_p/F_{\rm p,max}$ vs. $h=H_a/H_{\rm irr}$-plot can be calculated directly to $h_0=$ 0.5.
For a cross check of the data, it was necessary to extract the $H_{\rm irr}$-data from their paper.
However, the data published stem from a transport measurement employing pulsed currents,
whereas the $F_p$-data were obtained from a SQUID magnetometer. Luckily enough, $H_{\rm irr}(T)$
can be extracted from their $F_p(H)$-graphs (their figure D2).
According to the authors, the possible pinning centers of the $\delta T_c$-type may originate
in the local phase variation due to the oxygen or/and fluorine inhomogeneities.\\
The thin film data (3) \cite{37} exhibit a peak position at 0.33, and the published data were shown to
fit excellently to the Kramer theory, which is similar to the findings of other thin film materials.\\
The authors of (4) \cite{38} had plotted their data again in a scaling versus $H(F_{\rm p,max})$, but did
not perform any fits to the theory. Here, the data could be reworked in the same way like that of
Ref. \cite{36}, allowing them to be included here. Also here the peak position in the $F_{\rm p}$-scaling
of 0.42 indicated flux pinning provided by $\delta T_c$-type pinning being active in the samples.\\
The data (5) are exceptional in the sense that
these authors \cite{39} calculated $F_p$-data and showed in their paper a kind of scaling but
only for one temperature (20 K), however, for various Y-doping concentrations.
Nevertheless, a fit to these data provided a very high peak position of 0.71, which is the
highest value measured so far. Also, these materials, although
being polycrystalline, exhibit a very large irreversibility field which is just outside the
experimental field range of 16 T. Besides the fluorine substitution at oxygen site,
the substitution of Nd$^{3+}$ with a relatively smaller ion like Y$^{3+}$ creates lattice defects
in NdFeAs$_{0.7}$F$_{0.3}$ and thereby improves the flux pinning capability of the system. Therefore,
one may state here that it is obviously possible to even further increase the
$\delta T_c$-type pinning contribution within the 1111-family, which already shows the highest
peak positions of all Fe-based superconductors.

\subsection{11-family}
The 11-type material may be an interesting one for applications as it does not contain a
toxic material like As and no expensive rare-earth material, so the production
costs could be lowered. Therefore, some investigations concerning the pinning
force scaling on this type of material can be found in the literature.
These data \cite{40,41,42,43} are summarized in Table 3 and Figure 3.
Figure 3 shows clearly that the peak positions of the pinning force scaling are all located
between 0.28 and 0.33, indicating a flux pinning being dominated by small normal-conducting
particles. Therefore, one can say that the flux pinning in the 11-family behaves similar to
pure YBCO material. Additionally, the pinning function of the data set (13) of the 122-family is drawn here for
comparison as this material is also a iron selenide material. Even though the peak position of this data set is
located at 0.32, the high-field side of the pinning function is different and more similar to the other 122-type
data.

\subsection{111-family}
In this type of material, only one report concerning the pinning force scaling
is reported so far. The work of Shlyk {\em et al.} \cite{44} reports a good scaling of $F_{\rm p}$ versus
the peak field, $H(F_{\rm p,max}$ in Ga-doped LiFeAs.
They obtained $p=$ 2.06  and $q=$ 0.71 from a fit to the Kramer theory. However, as the peak
position in the pinning force scaling diagram is calculated via $h_0=p/p+q$, this would
yield $h_0=$ 0.74. This would be unreasonably high, so the data set had to be completely
reworked. The irreversibility field, $H_{\rm irr}$, could be determined from their
Fig. 3 and the given functional dependence. Plotting
then the data in the form $F_{\rm p}/F_{\rm p,max}$ vs. $h$ yields $p=$ 3.73  and $q=$ 3.98
as shown in Table 4. The peak position determined here, $h_0$, is still larger than 0.33, so the
main conclusion of Shlyk {\em et al.} remains valid. This shows that the Ga-doping is
introducing a stronger variation of $T_c$ as compared to undoped samples of the same type. Additionally,
the crtical current densities of an undoped sample were measured in Ref. \cite{44}, but no $F_{\rm p}$-scaling
was given which would allow a comparison.

\subsection{Remarks and analysis of the pinning force data}

The present collection of pinning force data allows to draw some important conclusions about the
flux pinning behavior in the Fe-based superconductors.
\begin{itemize}
\item[(i)] the data of practically all thin films and IBAD materials show only small values of
$h_0 \approx$ 0.33, which reflects the specific pinning landscape of the films on a substrate.
These characteristic defects cause a flux pinning with $h_0 \approx$ 0.2 to 0.33. The high
density of substrate-induced defects yields a much stronger contribution to the flux pinning,
and follows the Kramer theory of planar pins. In this sense, the mostly weaker $\delta T_c$-pinning
contribution provided by the material disorder does not have an influence on the resulting flux
pinning behavior.

\item[(ii)] The 1111- and 122-families show a clear tendency towards the presence of a pinning contribution
by a local variation of the transition temperature. Such a spatial variation of the superconducting gap
parameter which corresponds to a spatial variation of the transition temperature, $T_c$, was
accordingly observed in low-temperature STM data of Ba(Fe$_{0.9}$Co$_{0.1}$)$_2$As$_2$
in the review \cite{Hoffman} and in Refs. \cite{Yin,Massee}.

\item[(iii)] The strongest contribution of the $\delta T_c$-type pinning is observed in the 1111-family,
where the replacement of Nd by Y leads to the formation of strong lattice defects providing a strong
$\delta T_c$-pinning contribution. Furthermore, the variation of the peak position, $h_0$, is
extremely strong; there are not two data sets which yield approximately the same $h_0$-values. This is a
clear indication that more characterization experiments are needed in this family.
The observation of Ref. \cite{39} indicates that one may find still more materials with a strong
flux pinning.

\item[(iv)] The 122-type family consists of iron pnictides and selenides, but the overall behavior of the
flux pinning scaling is similar, with a tendency towards the $\delta T_c$-contribution, which is especially
visible on the high-field side of the scaling graph.

\item[(v)] The flux pinning data of the 11-family obtained in the literature are strikingly similar to
each other, yielding peak positions between 0.27 and 0.32. These materials behave therefore very
similar to pure Y-123 material, and the flux pinning is of the collective type.

\item[(vi)] In the 111-family, there is currently only one report on flux pinning, but these data
indicate that doping strongly increases the $\delta T_c$-type pinning contribution.

\end{itemize}

The flux pinning properties of the Fe-based superconductors show a large variety of behaviors as illustrated
in Figs. 1 to 4 and Tables 1 to 4. This is also reflected in the variation of the scaling parameters $p$ and $q$,
which often deviate from the DH values of 1 and 2 for the normal point pinning and the $\delta T_c$-type point pinning.
The values for $p$ range between 0.23 and 3.73, the ones of $q$ between 0.33 and 4.7. Values larger than 2 do not exist
in the model of DH. However, here it is important to note that the relation between $p$ and $q$, which is manifested 
by the peak position $h_0$, is always in a reasonable range.
In many cases, the fitting curves show deviations especially at the high-field side of the scaling diagram, which 
can be explained by flux creep effects. Such a behavior was already discussed in Ref. \cite{MK_PRB}, and this may also
lead to sets of $p$ and $q$ being different from the DH model.  

The 122- and 1111-families exhibit a clear tendency towards a strong contribution of the $\delta T_c$-pinning, which is manifested by the peak positions of the scaling diagrams being larger than 0.4. The work of Ref. \cite{39} demonstrated that by chemical doping this contribution can be stregthened, so their material exhibits the highest peak position measured so far in any type of high-$T_c$ superconductor. Therefore, the chemical
doping enables to tune the flux pinning behavior also in most of the iron pnictide and selenide compounds.

The data available in the literature for the 1111-family are largely scattered as not two reports yield the same peak
position of the pinning force scaling. Therefore, it is difficult to give conclusions here, but the strong $\delta T_c$-pinning contribution is already evident. Further work is needed here to fully understand the flux pinning behavior in these materials. 
The chemical doping of the superconducting compounds leads obviously to a strong spatial variation of the superconducting properties,
which may create a situation similar to the ternary 123-type compounds \cite{14,MK_APL,MK_EPL}. 
However, detailed measurements 
by, e.g., low-temperature STM and $m(T,B)$ magnetization field-cooling curves is still lacking in the literature.

From the 122-type iron selenides and the 111-family, there are still only one data set of each published in the literature. 
Recent reports like Ref.\cite{Ying} show that different chemical compositions also yield different shapes of the 
magnetization loops, which also indicates that the flux pinning behavior of the 122-iron selenides will also exhibit 
a large variety.

Concerning possible applications of the iron-based superconducting materials, one can state that the flux pinning
scaling behavior is in all cases strongly different from that of MgB$_2$, which has a similar value of the transition temperature.
Therefore, this will enable high-field applications of the iron-based superconducting materials, as their behavior is more 
similar to that of the cuprate high-$T_c$ superconductors.

\section{Conclusions}
In this contribution, the literature data of the flux pinning force scaling in iron pnictides and selenides
are summarized. The data obtained by many authors could be included directly in the analysis, while some data sets scaled by
the field of the pinning force maximum had to be reworked and replotted. Finally, the results of 30 different measurements 
are compared to each other. From this, some important conclusions concerning the further development of these materials towards
possible applications can be drawn. The 122- and 1111-families exhibit a clear tendency towards a strong $\delta T_c$-pinning
contribution to the flux pinning; the data of Y-doped NdFeAs$_{0.7}$F$_{0.3}$ exhibit the highest peak position of the 
pinning force scaling measured on any high-$T_c$ superconducting material. In contrast to this, the 11-family shows a 
typical collective flux pinning. Of the 111-family, only one experiment exists in the literature, but also here the chemical
doping seems to introduce the $\delta T_c$-pinning contribution.

\begin{acknowledgments}
One of the authors (M.R.K) thanks D. Johrendt and E. Wiesenmayer (LMU M\"unchen) for the polycrystalline 122-sample.

\end{acknowledgments}

\newpage

{\large\bf Figure captions:}\\[0.2cm]

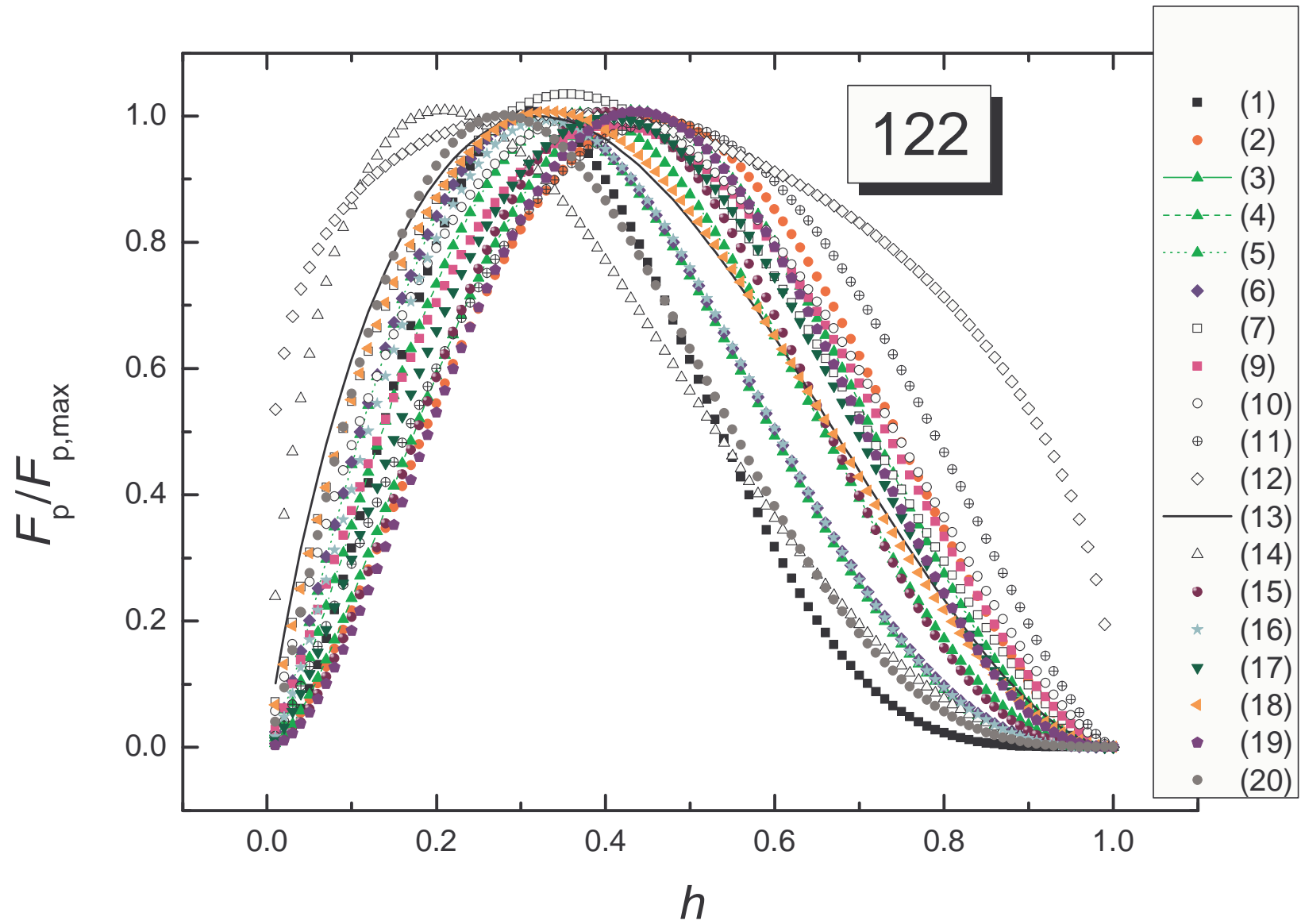
\begin{figure}[h]
\caption[]{Pinning force scaling, $F_{\rm p}/F_{\rm p,max}$ versus $h=H_a/H_{\rm irr}$, for the 122-family -- both
   pnictides and selenides. The open symbols are used for thin film (7, 10, 11) and IBAD tape (12, 14) data, the filled 
   symbols denote    single crystal data, and the one data set of a polycrystalline sample (16) is indicated by stars. The data set
   (13) is the only experiment of a 122-selenide material (indicated by a darker background), and is drawn using a solid black line. Except for the
   thin film/IBAD tape data, a tendency towards peak positions $h_0 >$ 0.33 is evident from the curves presented.}
\label{F1}
\end{figure}

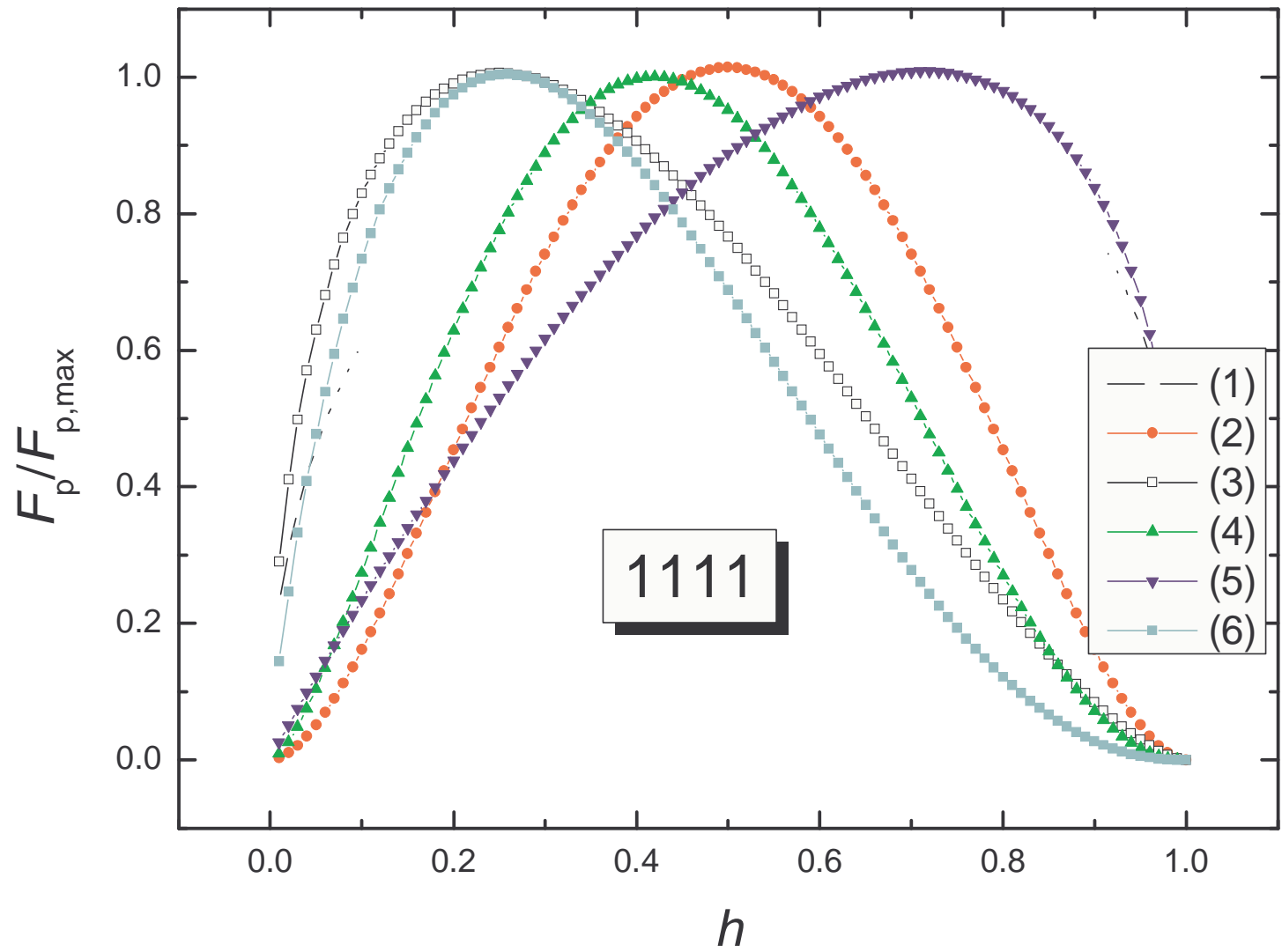
\begin{figure}[h]
\caption[]{Pinning force scaling, $F_{\rm p}/F_{\rm p,max}$ versus $h=H_a/H_{\rm irr}$, for the 1111-family.
    The open symbols are used for the thin film data (3), polycrystalline materials and single crystals are indicated using 
    full symbols. The peak position of the thin film data is the lowest of all; all other curves exhibit quite
    high but distinctly different $h_0$-values, with (5) yielding the largest $h_0$ of all iron-based superconducting materials investigated so far.}
\label{F2}
\end{figure}

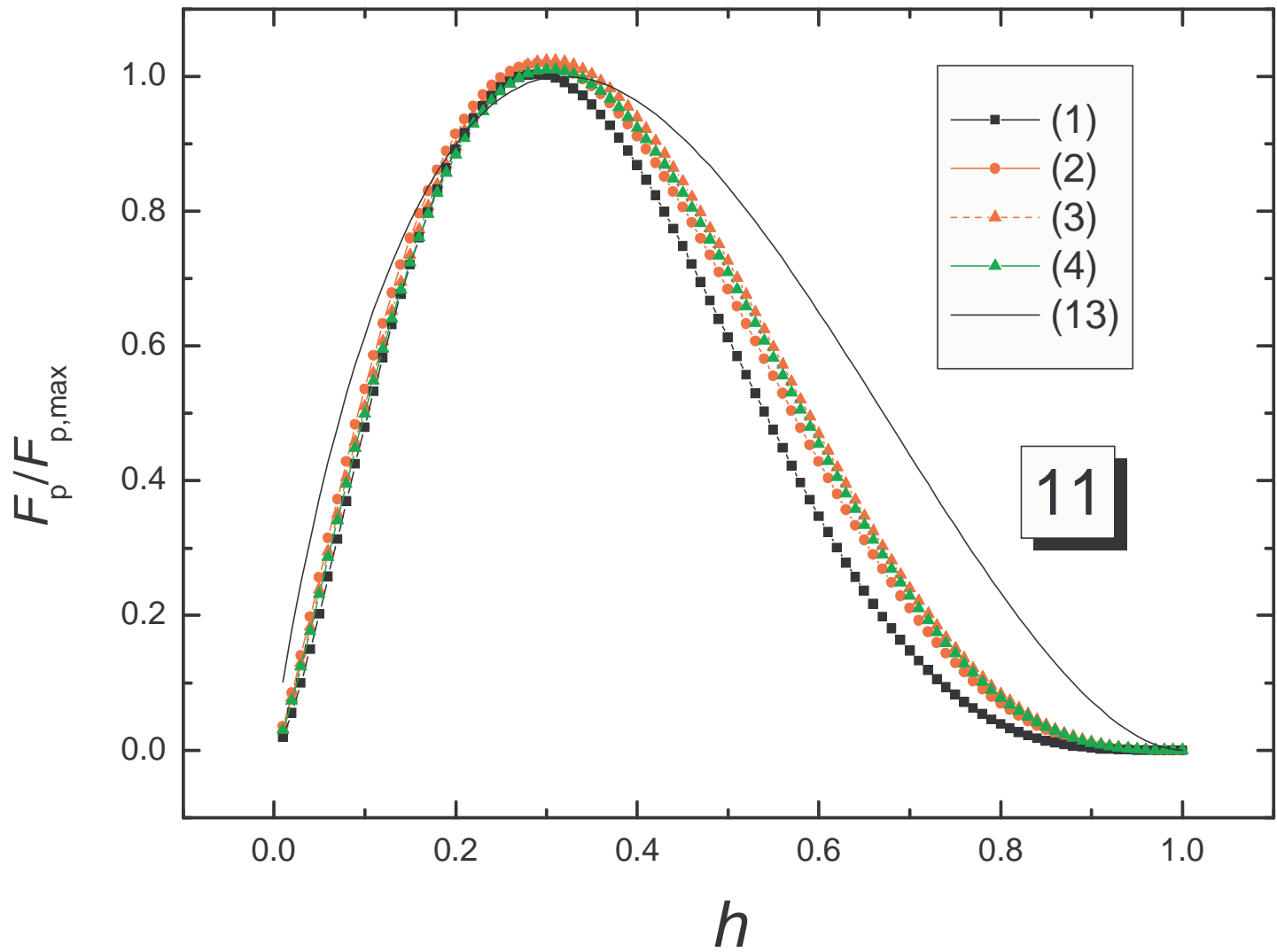
\begin{figure}[h]
\caption[]{Pinning force scaling, $F_{\rm p}/F_{\rm p,max}$ versus $h=H_a/H_{\rm irr}$, for the 11-family. All
     investigated materials are single crystals in this case. Practically all published data fall together, and the resulting peak position, $h_0$, is approx. 0.3. For comparison, the data set (13) of the 122-family, which is also a iron selenide, is added for comparison. Although the peak position is similar to the other 11-family data, the
     high-field side reveals a much stronger pinning contribution.}
\label{F3}
\end{figure}

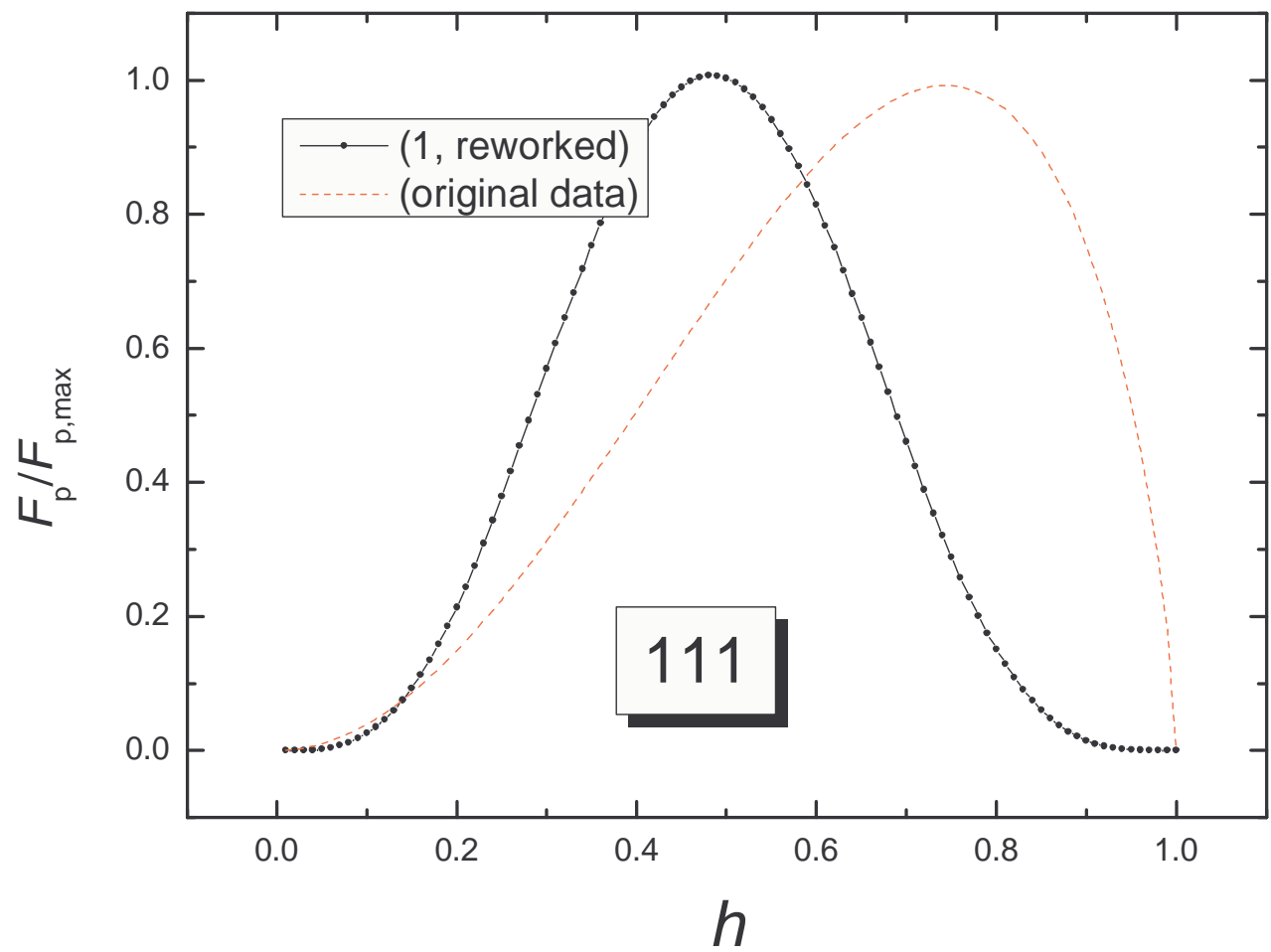
\begin{figure}[h]
\caption[]{Pinning force scaling, $F_{\rm p}/F_{\rm p,max}$ versus $h=H_a/H_{\rm irr}$, for the 111-family. For this
    material, only one data set is published in the literature. The dashed line indicates the scaling function when using the fit parameters as published by the original authors; the symbols give the pinning function after a
    complete reworking of the data. The relative high peak position persists after the necessary reworking.}
\label{F4}
\end{figure}

\end{document}